\input{aipcheck}

\documentclass[
    ,final            
  ]
  {aipproc}

\layoutstyle{6x9}

\begin{document}

\title{First Measurement Of The Jet Cross Section In Polarized p+p Collisions At $\sqrt{s}=200 $ GeV}


\author{Michael L. Miller (for the STAR Collaboration)}{
  address={Massachusetts Institute of Technology, Department of Physics,  77 Massachusetts Avenue, Cambridge, MA  02139 }
}

\begin{abstract}
We present preliminary measurements of the inclusive jet cross
section in the transverse momentum region 5<$p_{T}$<50
GeV/\textit{c} from
0.2 $pb^{-1}$ of polarized p+p data at $\sqrt{s}$=200 GeV.
The data were corrected for detector inefficiency and resolution
using PYTHIA events processed through a full GEANT simulation.  The
measured jet cross section agrees well with NLO pQCD calculations
over seven orders of magnitude.  These "proof of principal"
measurements pave the way for ongoing analyses of the higher
statistics ($\sim$ 3 $pb^{-1}$) data sample from the 2005 RHIC run.
\end{abstract}

\maketitle


The relativistic heavy ion collider (RHIC) is the first high energy
polarized proton-proton collider.  Of the two large detectors, the
solenoidal tracker at RHIC (STAR) is uniquely capable of full jet
reconstruction in p+p collisions.  STAR is just embarking on a rich
and diverse set of jet measurements in p+p collisions.  The
calibration of the jet energy scale and transverse energy resolution
are critical to many physics analyses including, but not limited to,
constraint of the gluon polarization via measurement of jet
production in polarized p+p collisions \cite{joanna}.  Future high
statistics measurement of unpolarized jet cross sections at RHIC may
provide significant constraints to previously measured large-x
parton distribution functions (PDFs), which could impact searches
for new physics at both the Tevatron and the LHC \cite{pumplin}.  We
present a preliminary measurement of the inclusive jet cross section
for 5<$p_T<$50 GeV/$c$. The measurement, primarily limited by
systematic uncertainties, agrees reasonably well with a NLO pQCD
calculation.


The data were collected during a short 2004 p+p commissioning run at
200 GeV.  STAR, a large acceptance collider detector with precision
tracking and electromagnetic calorimetry, is described in detail
elsewhere \cite{STAR-nim}.  An array of scintillating beam-beam
counters (BBC) spanning pseudorapidity 3.3<|$\eta$|<5 were used to
trigger on non-singly diffractive (NSD) inelastic collisions.  The
BBCs were sensitive to 26.1$\pm$2.0 mb (87\%) of the NSD cross
section \cite{Gans}. The detector subsystems of principle interest
to this analysis were the large acceptance time projection chamber
(TPC) and the partially commissioned (2400 of 4800 towers) barrel
electromagnetic calorimeter (BEMC).  Both have full azimuthal
coverage, and the TPC (BEMC) covers $| \eta |$<1.2 (0<$\eta$<1). The
BEMC was calibrated \textit{in situ} using the high statistics data
from the immediately preceding Au run. The relative gain of the BEMC
was established by studying hadronic energy deposition of charged
tracks, and the overall energy scale was set using well contained
showers from 1.5<$p$<8 GeV/\textit{c} electrons selected in the TPC.
The data presented here were collected in two trigger
configurations: minimum bias (MB) and high tower (HT). The highly
pre-scaled MB trigger required a BBC coincidence between positive
and negative $\eta$ in time with a valid bunch crossing. The HT
trigger required the MB condition as well as a single tower above an
$E_{T}$ threshold corresponding to $\sim$2.5 ($\sim$3.0) GeV at
$\eta_{tower}$ of 0 (1). This trigger was efficient for energetic
$\gamma$, $\pi^{0}$ and $e$ candidates. It also significantly
enhanced the population of high $p_T$ jets, but with strong
energy-dependant efficiency.  In this analysis we attempt to correct
for this jet-$p_T$ dependent trigger efficiency using simulation.
The produced 2004 event sample consisted of $\sim$4.2M events before
cuts, corresponding to a sampled luminosity of $\sim$0.16 $pb^{-1}$.


To preserve a large (80-90\%) and uniform tracking efficiency,
events with a primary vertex within z=$\pm$60 cm of the center of
STAR were analyzed.  Additionally, a requirement of
$E_{T}^{trig-tower}$>3.5 GeV was imposed on HT events to ensure
uniform trigger efficiency across all BEMC towers.  We analyzed
charged particles with $p_{T}$>0.2 GeV/\textit{c} originating within
a 3 cm sphere of the reconstructed primary interaction point. Tower
energies were corrected for hadronic energy deposition. Towers with
corrected $E_{T}$<0.2 GeV/\textit{c} were discarded from the
analysis. Jet finding was performed using the midpoint-cone
algorithm \cite{blazey}. For each event, a list of four momenta was
constructed from TPC tracks and BEMC towers.  For tracks (towers)
the $\pi^+$ ($\gamma$) mass was assumed.  The algorithm parameters
used were $r_{cone}$=0.4, $p_{T}^{seed}$=0.5 GeV/\textit{c}, and
$f_{merge}^{split}$=0.5. However, it should be noted that the HT
trigger requirement introduced an effective seed threshold of 3.5
GeV. Reconstructed jets were reported for $p_{T}$>5 GeV/\textit{c}
and were restricted to 0.2<$\eta$<0.8 to minimize edge effects in
the BEMC. Fake "neutral energy" jets originating from accelerator
backgrounds were discarded \cite{joanna}. After all cuts
approximately 10k (55k) jets remained from the MB (HT) sample. We
attempt to correct the measured jet yield to the particle level,
i.e. the energy of the $r_{cone}$=0.4 cluster immediately after
hadronization but before any detector effects.  
Detector efficiency and finite resolution were evaluated by studying
PYTHIA (v6.205) \cite{pythia} events passed through the STAR
simulation and reconstruction software.  The "true" properties of
jets were defined by running the same clustering algorithm on the
final state PYTHIA event record. The "accepted" properties of jets
were defined by clustering the simulated track and tower output of
the full simulation/reconstruction chain.  The jet $p_T$ resolution
was found to be 25\%, which motivated the final choice of binning.
We define a "generalized efficiency" via the bin-by-bin correction
factor $c(p_T) =
\frac{M^{acc}(p_{T}^{acc})}{N^{true}(p_{T}^{true})}$
%
%
that convolutes effects from jet reconstruction efficiency,
resolution induced bin sharing, and trigger efficiency. The
correction factor is evaluated separately for MB and HT modes, where
the Monte Carlo events are subject to the same trigger requirements
as the data.  Note that bin-migration can yield $c>1$.

Figure 1a shows the correction factor for MB and HT data.  Whereas
the MB correction is $\approx 1\pm0.1$, the HT correction varies by
two orders of magnitude.  The primary difference between the MB and
HT corrections is the inclusion of the HT trigger efficiency, which
varies strongly with jet-$p_T$ due to the weak correlation between
$p_T$ of the trigger photon (via $\pi^0$ decay) and $p_T$ of the
parent jet. Figure 1b shows the corrected jet cross section. An
additional correction of $60\pm$2\% was applied to account for valid
MB events where no primary vertex was reconstructed.  The MB and HT
data show good agreement in the three overlapping bins, where the MB
and HT efficiencies differ by two orders of magnitude.  The data are
compared to results from a fast (small-cone approximation) NLO pQCD
calculation incorporating CTEQ6M PDFs and $r_{cone}$=0.4, with a
scale of $\mu_F~=~\mu_R~=~p_T$ \cite{werner}.  It was verified that
the calculation agrees with the standard EKS NLO calculation
\cite{eks} to better than 1\%. The calculations were performed with
the same binning as the data and both data and calculation are
plotted in the middle of the bins. We
\begin{tabular}{p{7cm}c}
  \textbf{Figure 1:} \begin{small}(a) The correction factor for MB and HT data
  derived from simulation.  Statistical errors are shown.  (b)
  Preliminary inclusive jet cross section compared to NLO pQCD
  calculation.  Statistical uncertainties are plotted but are
  smaller than the markers.  (c) Ratio comparison of data vs.
  theory.  The shaded band represents the dominant systematic
  uncertainty, and an 8\% uncertainty on the overall normalization is
  not shown.  See text for details. \end{small} &
  \raisebox{-7cm}{
  \includegraphics[clip=true, viewport=0.07in 0.in 6.6in 7.5in, height=.37\textheight]{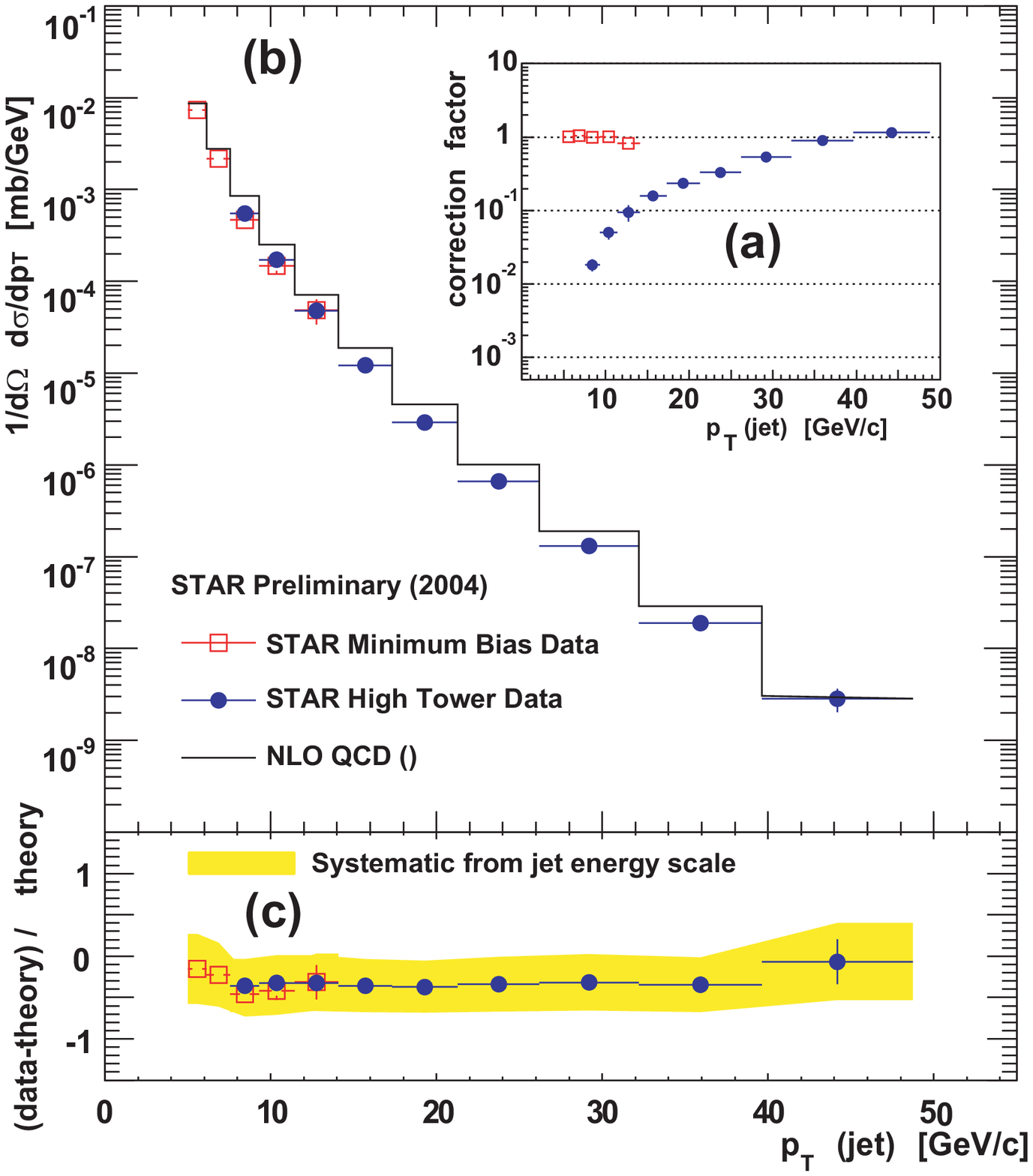}
  }
\end{tabular}
find good shape agreement over seven orders of magnitude.  Figure 1c
shows the ratio between data and theory.  The shaded band represents
the dominant systematic uncertainty (50\% change in yield) on the
measured cross section from the 10\% uncertainty on the jet energy
scale.  For $p_T>10$ GeV/c we find a systematic offset between data
and theory, but we conclude reasonable agreement within the large
systematic uncertainty.  Given the steeper slope of the jet cross
section at RHIC energies we expect the cross section measurement to
show increased sensitivity to uncertainties in the jet energy scale
when compared to the Tevatron.  A major challenge for future RHIC
runs will be the use of smaller cross section processes such as
di-jet and photon-jet final states to hopefully reduce the
uncertainty on the jet energy scale to below 5\%.


We have presented preliminary measurement of the inclusive jet cross
section from 0.16 $pb^{-1}$ of polarized \textit{p+p} collisions at
$\sqrt{s}=200$ GeV.
We correct for a high tower trigger efficiency that changes by two
orders of magnitude over the range of the measurement.
The reasonable agreement with NLO pQCD calculations over seven
orders of magnitude motivates the application of perturbative QCD to
interpret the spin dependent jet asymmetries recently reported from
STAR.  These measurements pave the way for ongoing analyses of
3$pb^{-1}$ of data already collected.



\bibliographystyle{aipproc}   

\end{document}